\documentclass{aa}
\usepackage{euclid}
\usepackage{graphicx}
\usepackage{natbib}
\bibpunct{(}{)}{;}{a}{}{,} 
\usepackage[varg]{txfonts}

\newcommand{\nc}{\newcommand}

 \nc{\br}{{\bf r}}
 \nc{\bx}{{\bf x}}
 \nc{\half}{{\textstyle \frac12}}
 \nc{\quarter}{{\textstyle \frac14}}
 \nc{\bear}{\begin{eqnarray}} \nc{\eear}{\end{eqnarray}}
 \nc{\beq}{\begin{equation}} \nc{\eeq}{\end{equation}}
 \nc{\nn}{\nonumber}
 \nc{\farc}{\frac}
 \nc{\wxi}{\hat{\xi}}
 \nc{\wxiLS}{\hat{\xi}_\mathrm{LS}}
 \nc{\rmax}{r_\mathrm{max}}
 
 \nc{\Nd}{N_{\rm d}}
 \nc{\Nr}{N_{\rm r}}
 \nc{\Nrp}{N'_{\rm r}}
 \nc{\Ms}{M_{\rm s}}
 \nc{\Mr}{M_{\rm r}}
 \nc{\Md}{M_{\rm d}}
 \nc{\Gp}{G^{\rm p}}
 \nc{\Gt}{G^{\rm t}}
 \nc{\tr}{t_{\rm r}}
 \nc{\pr}{p_{\rm r}}
 \nc{\pc}{p_{\rm c}}
 \nc{\qr}{q_{\rm r}}
 \nc{\tur}{t^{\rm r}}
 \nc{\pur}{p^{\rm r}}
 \nc{\puc}{p^{\rm c}}
 \nc{\qur}{q^{\rm r}}
 \nc{\trp}{t'_{\rm r}}
 \nc{\prp}{p'_{\rm r}}
 \nc{\prpp}{p''_{\rm r}}
 \nc{\turp}{t^{{\rm r}'}}
 \nc{\purp}{p^{{\rm r}'}}
 \nc{\purpp}{p^{{\rm r}''}}
 
\nc{\ek}[1]{\textcolor{blue}{#1}}
\nc{\va}[1]{\textcolor{gray}{#1}}
\nc{\fma}[1]{\textcolor{cyan}{#1}}
\nc{\hxc}[1]{\textcolor{purple}{#1}}
\nc{\eb}[1]{\textcolor{black}{#1}}
\nc{\hx}[1]{\textcolor{black}{#1}}
 
\usepackage{color}

\begin{document} 

\title{Estimating the galaxy two-point correlation function using a split random catalog}

\author{E. Keih\"{a}nen\inst{\ref{inst1}}\and H. Kurki-Suonio\inst{\ref{inst1}}\and V. Lindholm\inst{\ref{inst1}}\and A. Viitanen\inst{\ref{inst1}}\and A.-S. Suur-Uski\inst{\ref{inst1}}\and 
V. Allevato\inst{\ref{inst2},\ref{inst1}}\and E. Branchini\inst{\ref{inst3}}\and F. Marulli\inst{\ref{inst4},\ref{inst5},\ref{inst6}}\and P. Norberg\inst{\ref{inst7}}\and D. Tavagnacco\inst{\ref{inst8}}\and S. de la Torre\inst{\ref{inst9}}\and J. Valiviita\inst{\ref{inst1}}\and M. Viel\inst{\ref{inst10},\ref{inst11},\ref{inst8},\ref{inst14}}\and J. Bel\inst{\ref{inst12}}\and M. Frailis\inst{\ref{inst8}}\and A. G. S\'anchez\inst{\ref{inst13}}}

\institute{Department of Physics and Helsinki Institute of Physics, Gustaf H\"{a}llstr\"{o}min katu 2, 00014 University of Helsinki, Finland\label{inst1}
\and Scuola Normale Superiore, Piazza dei Cavalieri 7, I-56126 Pisa, Italy\label{inst2}
\and Department of Mathematics and Physics, Roma Tre University, Via della Vasca Navale 84, I-00146 Rome, Italy\label{inst3}
\and Dipartimento di Fisica e Astronomia - Alma Mater Studiorum Universit\`{a} di Bologna, via Piero Gobetti 93/2, I-40129 Bologna, Italy\label{inst4}
\and INAF - Osservatorio di Astrofisica e Scienza dello Spazio di Bologna, via Piero Gobetti 93/3, I-40129 Bologna, Italy\label{inst5}
\and INFN - Sezione di Bologna, viale Berti Pichat 6/2, I-40127 Bologna, Italy\label{inst6}
\and ICC \& CEA, Department of Physics, Durham University, South Road, DH1 3LE UK\label{inst7}
\and INAF, Osservatorio Astronomico di Trieste, via Tiepolo 11, I-34131 Trieste, Italy \label{inst8}
\and Aix Marseille Univ, CNRS, CNES, LAM, Marseille, France\label{inst9}
\and SISSA, International School for Advanced Studies, Via Bonomea 265, 34136 Trieste TS, Italy \label{inst10}
\and INFN, Sezione di Trieste, Via Valerio 2, 34127 Trieste TS, Italy \label{inst11}
\and Aix Marseille Univ, Universit\'e de Toulon, CNRS, CPT, Marseille, France \label{inst12}
\and Max-Planck-Institut f\"ur extraterrestrische Physik, Postfach 1312, Giessenbachstr., 85741 Garching, Germany \label{inst13}
\and IFPU - Institute for Fundamental Physics of the Universe, Via Beirut 2, 34014 Trieste, Italy \label{inst14}
}

\date{Received XXX / Accepted YYY}

\abstract{The two-point correlation function of the galaxy distribution is a key
cosmological observable that allows us to constrain the dynamical and
geometrical state of our Universe. To measure the correlation function we
need to know both the galaxy positions and the expected galaxy density
field. The expected field is commonly specified using a Monte-Carlo
sampling of the volume covered by the survey and, to minimize additional
sampling errors, this random catalog has to be much larger than the data
catalog. Correlation function estimators compare data--data pair counts to
data--random and random--random pair counts, where random--random pairs
usually dominate the computational cost. Future redshift surveys will
deliver spectroscopic catalogs of tens of millions of galaxies. Given the large number of random
objects required to guarantee sub-percent accuracy, it is of paramount
importance to improve the efficiency of the algorithm without degrading its
precision. We show both analytically and numerically that splitting the
random catalog into a number of subcatalogs of the same size as the data
catalog when calculating random--random pairs and excluding pairs across
different subcatalogs provides the optimal error at fixed computational
cost. For a random catalog fifty times larger than the data catalog, this
reduces the computation time by a factor of more than ten without
affecting estimator variance or bias.}

\keywords{large-scale structure of Universe - Cosmology: observations - Methods: statistical - Methods: data analysis}

\titlerunning{Two-point correlation function with split random catalog}
\authorrunning{E. Keih\"{a}nen at al.}

\maketitle

\section{Introduction}

The spatial distribution of luminous matter in the Universe is a key diagnostic for studying cosmological models and the physical processes involved in the assembly of structure. In particular, light from galaxies is a robust tracer of the overall matter distribution, whose statistical properties can be predicted by cosmological models. 
Two-point correlation statistics are very effective tools for compressing the cosmological information encoded in the spatial distribution of the mass in the Universe. In particular, the two-point correlation function in configuration space has emerged as one of the most popular cosmological probes. Its success stems from the presence of characterized features that can be identified, measured, and effectively compared to theoretical models to extract clean cosmological information.

One such feature is baryon acoustic oscillations (BAOs), which imprint a characteristic scale in the two-point correlation function that can be used as a standard ruler. After the first detection in the two-point correlation function of SDSS DR3 and 2dFGRS galaxy catalogs \citep{eisenstein05,cole05}, the BAO signal was identified, with different degrees of statistical significance, and has since been used to constrain the expansion history of the Universe in many spectroscopic galaxy samples (see e.g., \citet{percival10} \citet{blake11}, \citet{beutler11}, \citet{anderson12}, \citet{anderson14}, \citet{ross15}, \citet{alam17}, \citet{ross17}, \citet{vargas18}, \citet{bautista18}, and \citet{ata18}).
Several of these studies did not focus on the BAO feature only but also analyzed the anisotropies in the two-point correlation function induced by the peculiar velocities \citep{kaiser87}, the so-called redshift space distortions (RSD), and by assigning cosmology-dependent distances to the observed redshifts (the 
\citet{ap79} test).  For RSD analyses, see also, for example, \citet{peacock01}, \citet{guzzo08}, \citet{beutler12}, \citet{reid12}, \citet{delatorre17}, \citet{pezzotta17}, \citet{zarrouk18}, \citet{hou18}, and \citet{ruggeri19}.

Methods to estimate the galaxy two-point correlation function (2PCF)\ $\xi(\br)$ from survey data are based on its definition as the excess probability of finding a galaxy pair.
One counts from the data ($D$) catalog the number $DD(\br)$ of pairs of galaxies with separation $\bx_2-\bx_1 \in \br$, where $\br$ is a bin of separation vectors, and compares it to the number of pairs $RR(\br)$ in a corresponding randomly generated ($R$) catalog and to the number of data-random pairs $DR(\br)$.  
The bin may be a 1D ($r\pm\half\Delta r$), 2D, or a 3D bin.  In the 1D case, $r$ is the length of the separation vector and $\Delta r$ is the width of the bin.
From here on, `separation $\br$' indicates that the separation falls in this bin.   

Several estimators of the 2PCF have been proposed by \citet{hewett82}, \citet{dp83}, \citet{hamilton93}, and \citet{Landy:1993yu}, building on the original \citet{ph74} proposal.  
These correspond to different combinations of the $DD$, $DR$, and $RR$ counts to obtain a 2PCF estimate $\wxi(\br)$;  see \citet{Kerscher99} and \citet{Kerscher00} for more estimators.  The Landy--Szalay \citep{Landy:1993yu} estimator
 \beq
        \wxiLS(\br) \ := \ \frac{\Nr(\Nr-1)}{\Nd(\Nd-1)}\frac{DD(\br)}{RR(\br)} - \frac{\Nr-1}{\Nd}\frac{DR(\br)}{RR(\br)} + 1 
 \label{eq:LS}
 ,\eeq
(we call this method `standard LS' in the following) is the most commonly used, since it provides the minimum variance when $|\xi|\ll1$ 
and is unbiased in the limit $\Nr\rightarrow\infty$.  Here $\Nd$ is the size (number of objects) of the data catalog and $\Nr$ is the size of the random catalog.  We define $\Mr := \Nr/\Nd$.  To minimize random error from the random catalog, $\Mr \gg 1$ should be used  (for a different approach, see \citet{Demina2018}).

One is usually interested in $\xi(\br)$ only up to some $\rmax \ll L_\mathrm{max}$ (the maximum separation in the survey), and therefore pairs with larger separations can be skipped.  
Efficient implementations of the LS estimator involve pre-ordering of the catalogs  through kd-tree, chain-mesh, or other algorithms \citep[e.g.,][]{moore01,2012arXiv1210.1833A, 2015ascl.soft08007J, 2016AC....14...35M}
to facilitate this.
The computational cost is then roughly proportional to the actual number of pairs with separation $r \leq \rmax$.

The correlation function is small for large separations, and in cosmological surveys $\rmax$ is large enough so that for most pairs $|\xi(\br)| \ll 1$.
The fraction $f$ of $DD$ pairs with $r\leq\rmax$ is therefore not very different from the fraction of $DR$ or $RR$ pairs with $r\leq\rmax$.  The computational cost is dominated by the part proportional to the total number of pairs needed,
$\half f\Nd(\Nd\! -\! 1) + f\Nd\Nr + \half f\Nr(\Nr\! -\! 1) \approx \half f\Nd^2(1 + 2\Mr + \Mr^2)$,
which in turn is dominated by the $RR$ pairs as $\Mr \gg 1$.  The smaller number of $DR$ pairs contribute much more to the error of the estimate than the large number of $RR$ pairs, whereas the cost is dominated by $RR$.  Thus, a significant saving of computation time with an insignificant loss of accuracy may be achieved by counting only a subset of $RR$ pairs, while still counting the full set (up to $\rmax$) of $DR$ pairs.

A good way to achieve this is to use many  small   (i.e, low-density) $R$ catalogs instead of one large (high-density) catalog \citep{Landy:1993yu, WallJenkins, SlepianEisenstein}, or, equivalently, to split an already generated large $R$ catalog into $\Ms$ small ones for the calculation of $RR$ pairs while using the full $R$ catalog for the $DR$ counts.  {This method has been used by some practitioners (e.g., \citet{zehavi18}\footnote{I.~Zehavi, private communication}), but this is usually not documented in the literature.} One might also consider obtaining a similar cost saving by diluting (subsampling) the $R$ catalog for $RR$ counts, but, as we show below, this is not a good idea.  We refer to these two cost-saving methods as `split' and `dilution'.

In this work we  theoretically\textbf{ }derive the additional covariance and bias due to the size and treatment of the $R$ catalog; test these predictions numerically with mock catalogs representative of next-generation datasets, such as the spectroscopic galaxy samples that will be obtained by the future Euclid satellite mission \citep{laureijs11}; and show that the `split' method, while reducing the computational cost by a large factor, retains the advantages of the LS estimator. 

We follow the approach of Landy \& Szalay (1993; hereafter,  LS93), but generalize it in a number of ways: In particular, since we focus on the effect of the random catalog, we do not work in the limit $\Mr\rightarrow\infty$.  Also, we calculate covariances, not just variances, and make fewer approximations 
(see Sect.~\ref{sec:qta}).

The layout of the paper is as follows. In Sect.~\ref{sec:theory} we derive theoretical results for bias and covariance.  In Sect.~\ref{sec:split} we focus on the split LS estimator and its optimization.  In Sect.~\ref{sec:mocks} we test the different estimators with mock catalogs.  Finally, we discuss the results and present our conclusions in Sect.~\ref{sec:conclusions}.

\section{Theoretical results: bias and covariance}
\label{sec:theory}

\subsection{General derivation}

We follow the derivation and notations in {LS93} but extend to the case that includes random counts covariance.  We consider the survey volume as divided into $K$ microcells (very small subvolumes) and work in the limit $K\rightarrow\infty$, which means that no two objects will ever be located within the same microcell.

Here, $\alpha$, $\beta$, and $\gamma$  represent the relative deviation of the $DD(\br)$, $DR(\br)$, and $RR(\br)$ counts from their expectation values (mean values over an infinite number of independent realizations):
 \bear
        DD(\br) & =: & \langle DD(\br) \rangle [1+\alpha(\br)] \,,\nn\\
        DR(\br) & =: & \langle DR(\br) \rangle [1+\beta(\br)] \,,\nn\\
        RR(\br) & =: & \langle RR(\br) \rangle [1+\gamma(\br)] \,.
 \label{eq:deviations}
 \eear
{By definition $\langle\alpha\rangle=\langle\beta\rangle=\langle\gamma\rangle=0$.  The factors $\alpha$, $\beta$, and $\gamma$ represent fluctuations in the pair counts, which arise as a result of a Poisson process.
As long as the mean pair counts per bin are large ($\gg1$) the relative fluctuations will be small.
We calculate up to second order in $\alpha$, $\beta$, and $\gamma$, and ignore the higher-order terms}
{ (in the limit $M_r\rightarrow\infty$, $\gamma \rightarrow 0$, so LS93 set $\gamma = 0$ at this point).}

The expectation values for the pair counts are:
 \bear
        \langle DD(\br)\rangle & = & \half \Nd(\Nd-1)\Gp(\br)[1+\xi(\br)] \,,\nn\\
        \langle DR(\br)\rangle & = & \Nd\Nr \Gp(\br) \,,\nn\\
        \langle RR(\br)\rangle & = & \half \Nr(\Nr-1)\Gp(\br) \,,
 \label{eq:pair_expect}
 \eear
where $\xi(\br)$ is the correlation function normalized to the actual number density of galaxies in the survey
and
 \beq
        \Gp(\br) \ := \ \farc{2}{K^2} \sum_{i<j}^K \Theta^{ij}(\br)
 \eeq
is the fraction of microcell pairs with separation $\br$.  Here $\Theta^{ij}(\br) := 1$ if $\bx_i-\bx_j$ falls in the $\br$-bin, otherwise it is equal to zero. 

The expectation value of the 
LS estimator \eqref{eq:LS} is 
 \bear
        \langle\wxiLS\rangle & = & (1+\xi)\left\langle\frac{1+\alpha}{1+\gamma}\right\rangle - 2\left\langle\farc{1+\beta}{1+\gamma}\right\rangle + 1 \nn\\
        & \eqsim & \xi + (\xi-1)\langle\gamma^2\rangle + 2\langle\beta\gamma\rangle  \label{eq:LSbias_abc} \,.
 \eear

 A finite $R$ catalog thus introduces a (small) bias.  { (In LS93, $\gamma = 0$, so the estimator is unbiased in this limit.)}
This expression is calculated to second order in $\alpha$, $\beta$, and $\gamma$ (we denote equality to second order by `$\eqsim$').  { Calculation to higher order 
is beyond the scope of this work.} Since data and random catalogs are independent, $\langle \alpha\gamma\rangle=0$. 

{We introduce shorthand notations $\langle\alpha_1\alpha_2\rangle$ for $\langle\alpha(\br_1)\alpha(\br_2)\rangle$, $\langle DD_1 DD_2\rangle$ for $\langle DD(\br_1) DD(\br_2)\rangle$, and similarly for other terms.}

For the covariance we get 
 \beq
 \begin{split}
        &\mbox{Cov}\left[\wxiLS(\br_1),\wxiLS(\br_2)\right] \ \equiv \ \left\langle\wxiLS(\br_1)\wxiLS(\br_2)\right\rangle - \left\langle\wxiLS(\br_1)\right\rangle\left\langle\wxiLS(\br_2)\right\rangle \\
        &\quad \eqsim \ (1+\xi_1)(1+\xi_2)\langle\alpha_1\alpha_2\rangle + 4\langle\beta_1\beta_2\rangle + (1-\xi_1)(1-\xi_2)\langle\gamma_1\gamma_2\rangle  \\
        &\qquad - 2(1+\xi_1)\langle\alpha_1\beta_2\rangle - 2(1+\xi_2)\langle\beta_1\alpha_2\rangle \\
        &\qquad - 2(1-\xi_1)\langle\gamma_1\beta_2\rangle - 2(1-\xi_2)\langle\beta_1\gamma_2\rangle    \,.  
 \end{split}
 \label{eq:LScov_abc}
 \eeq   
Terms with $\gamma$ represent additional variance due to finite $\Nr$, and are new compared to those of{ LS93}.  Also, $\langle\beta_1\beta_2\rangle$ collects an additional contribution, { which we denote by $\Delta\langle\beta_1\beta_2\rangle$,} from variations in the random field { (see Sect.~\ref{sec:ptq})}.
The cross terms $\alpha_1\beta_2$ and $\alpha_2\beta_1$, instead depend linearly on the random field,
and average to the $\Nr\rightarrow\infty$ result.
The additional contribution  due to finite $\Nr$ is thus
 \beq
 \begin{split}
        &\Delta\mbox{Cov}\left[\wxiLS(\br_1),\wxiLS(\br_2)\right] \ \eqsim \
        4\Delta\langle\beta_1\beta_2\rangle + (1-\xi_1)(1-\xi_2)\langle\gamma_1\gamma_2\rangle \\
        &\qquad - 2(1-\xi_1)\langle\gamma_1\beta_2\rangle - 2(1-\xi_2)\langle\beta_1\gamma_2\rangle \,.
 \end{split}
 \label{eq:LSdeltacov}
 \eeq 

{ From \eqref{eq:deviations}, 
 \beq
        \langle DD_1\ DD_2 \rangle  \ = \ \langle DD_1\rangle\langle DD_2\rangle\left(1+\langle\alpha_1\alpha_2\rangle\right)
 ,\eeq
and so on, so that the covariances of the deviations are obtained from}
 \bear
        \langle\alpha_1\alpha_2\rangle & = & \farc{\langle DD_1\ DD_2 \rangle - \langle DD_1\rangle\langle DD_2\rangle}{\langle DD_1\rangle\langle DD_2\rangle} \,,\nn\\
        \langle\beta_1\beta_2\rangle & = & \farc{\langle DR_1\ DR_2 \rangle - \langle DR_1\rangle\langle DR_2\rangle}{\langle DR_1\rangle\langle DR_2\rangle} \,,\nn\\
        \langle\gamma_1\gamma_2\rangle & = & \farc{\langle RR_1\ RR_2 \rangle - \langle RR_1\rangle\langle RR_2\rangle}{\langle RR_1\rangle\langle RR_2\rangle} \,,\nn\\
        \langle\alpha_1\beta_2\rangle & = &\farc{\langle DD_1\ DR_2 \rangle - \langle DD_1 \rangle\langle DR_2 \rangle}{\langle DD_1 \rangle\langle DR_2 \rangle} \,,\nn\\
        \langle\beta_1\gamma_2\rangle & = & \farc{\langle DR_1\ RR_2 \rangle - \langle DR_1 \rangle\langle RR_2 \rangle}{\langle DR_1 \rangle\langle RR_2 \rangle} \,,\nn\\
        \langle\alpha_1\gamma_2\rangle & = & \farc{\langle DD_1\ RR_2 \rangle - \langle DD_1 \rangle\langle RR_1 \rangle}{\langle DD_1 \rangle\langle RR_2 \rangle} \ = \ 0\,.
 \label{deviations_cov}
 \eear

\subsection{Quadruplets, triplets, and approximations}
\label{sec:qta}

We use
 \beq
        \Gt_{12} := \Gt(\br_1,\br_2)  \ := \ \frac{1}{K^3}\sum^\ast_{ijk}\Theta^{ik}_1\Theta^{jk}_2
 \eeq
to denote the fraction of ordered microcell triplets, where $\bx_i-\bx_k \in \br_1$ and $\bx_j-\bx_k \in \br_2$.  
\hx{The notation $\sum^\ast$ means that only terms where all indices (microcells) are different are included.}
Here $\Gt_{12}$ is of the same magnitude as $\Gp_1\Gp_2$ but is larger.  

Appendix A gives examples of how the $\langle DD_1\ DD_2 \rangle$ and so on~in \eqref{deviations_cov} are calculated.
These covariances involve expectation values $\langle n_in_jn_ln_k\rangle$, where $n_i$ is the number of objects (0 or 1) in microcell $i$ and so on, and only cases where the four microcells are separated pairwise by $\br_1$ and $\br_2$ are included.  If all four microcells $i$, $j$, $k$, and $l$ are different, we call this case a quadruplet; it consists of two pairs with separations $\br_1$ and $\br_2$.  If two of the indices, that is, microcells, are equal, we have a triplet with a center cell (the equal indices) and two end cells separated from the center by $\br_1$ and $\br_2$.

We make the following three approximations:
\begin{enumerate}
\item For microcell quadruplets, the correlations between unconnected cells are approximated by zero on average. 
\item Three-point correlations vanish. 
\item The part of four-point correlations that does not arise from the two-point correlations vanishes.  
\end{enumerate}

With approximations (2) and (3), we have for the expectation value of a galaxy triplet
\beq
 \langle n_in_jn_k\rangle \propto 1+\xi_{ij}+\xi_{jk}+\xi_{ik} \,,
 \eeq
where $\xi_{ij} := \xi(\bx_j-\bx_i)$, and for a quadruplet
 \beq
 \langle n_in_jn_kn_l\rangle \propto 1+\xi_{ij}+\xi_{jk}+\xi_{ik}+\xi_{il}+\xi_{jl}+\xi_{kl}+\xi_{ij}\xi_{kl}+\xi_{ik}\xi_{jl}+\xi_{il}\xi_{jk} \,.
 \label{eq:quad}
 \eeq

We use  `$\simeq$' to denote results based on these three approximations.  
Approximation (1) is good as long as  the survey size is large compared to $\rmax$.  It allows us to drop terms other than $1+\xi_{ij}+\xi_{kl}+\xi_{ij}\xi_{kl}$ in \eqref{eq:quad}.  
Approximations (2) and (3) hold for Gaussian density fluctuations,
but in the realistic cosmological situation they are not good: the presence of the higher-order correlations makes the estimation of the covariance of $\xi(\br)$ estimators a difficult problem.  However, this difficulty applies only to the contribution of the data to the covariance,{ that is, to the part that does not depend on the size and treatment of the random catalog}.  The key point in this work is that while our theoretical result for the total covariance does not hold in a realistic situation (it is an underestimate), our results for {the difference in estimator covariance due to different treatments of the random catalog} hold well.

In addition to working in the limit $\Nr\rightarrow\infty$ ($\gamma = 0$), { LS93} considered only 1D bins and the case where $r_1 = r_2 \equiv r$ (i.e., variances, not covariances) and made also a fourth approximation: for triplets (which in this case have legs of equal length) they approximated the correlation between the end cells (whose separation in this case varies between $0$ and $2r$) by $\xi(r)$.   We use  $\xi_{12}$ to denote the mean value of the correlation between triplet end cells (separated from the triplet center by $\br_1$ and $\br_2$).  (For our plots in Sect.~\ref{sec:mocks} we make a similar approximation of  $\xi_{12}$ as Landy\&Szalay, { see Sect.~\ref{sec:var_bias}}.)  Also, { LS93} only calculated to first order in $\xi$, whereas we do not make this approximation.
\citet{bernstein94}  also considered covariances, and included the effect of three-point and four-point correlations, but worked in the limit $\Nr\rightarrow\infty$ ($\gamma = 0$).

\subsection{Poisson, edge, and $q$ terms}
\label{sec:ptq}

{ After calculating all the $\langle DD_1\ DD_2 \rangle$ and so on~(see Appendix \ref{sec:examples}), \eqref{deviations_cov} becomes}
\begin{gather}
  (1+\xi_1)(1+\xi_2)\langle\alpha_1\alpha_2\rangle \ \simeq \
   \frac{4}{\Nd}        
        (1+\xi_1)(1+\xi_2)\left[\frac{\Gt_{12}}{\Gp_1\Gp_2} - 1\right]  \nn  \\ 
  \quad + \frac{2(1+\xi_1)}{\Nd(\Nd-1)}\left[ \farc{\delta_{12}}{\Gp_1} - 2(1+\xi_2)\farc{\Gt_{12}}{\Gp_1\Gp_2} + (1+\xi_2) \right] \nn \\
  \quad + \frac{4(\Nd-2)}{\Nd(\Nd-1)}(\xi_{12}-\xi_1\xi_2)\farc{\Gt_{12}}{\Gp_1\Gp_2} 
 \,,\nn\\
   \langle\beta_1\beta_2\rangle \ \simeq \ \frac{1}{\Nd\Nr}\Biggl\{         
        \Nr\left[\frac{\Gt_{12}}{\Gp_1\Gp_2}-1\right] + \Nd\left[\frac{\Gt_{12}}{\Gp_1\Gp_2}-1\right]  \nn \\
  \qquad + 1 - \farc{2\Gt_{12}}{\Gp_1\Gp_2} + \farc{\delta_{12}}{\Gp_1} \Biggr\} 
     + \frac{\Nd-1}{\Nd\Nr}\xi_{12}\frac{\Gt_{12}}{\Gp_1\Gp_2}  
\,,\nn\\
 \langle\gamma_1\gamma_2\rangle \ = \ \frac{2}{\Nr(\Nr-1)}\left\{               
        2(\Nr-2)\left[\frac{\Gt_{12}}{\Gp_1\Gp_2} - 1\right] + \farc{\delta_{12}}{\Gp_1} - 1\right\} 
 \,,\nn\\
        \langle\alpha_1\beta_2\rangle \ \simeq \ \frac{2}{\Nd}\left[\frac{\Gt_{12}}{\Gp_1\Gp_2} - 1\right] \,,\nn\\     
        \langle\beta_1\gamma_2\rangle \ = \ \frac{2}{\Nr}\left[\frac{\Gt_{12}}{\Gp_1\Gp_2} - 1\right] \,,\nn\\
        \langle\alpha_1\gamma_2\rangle \ = \ 0,  
 \label{eq:alphabetaresults}
 \end{gather}
{ for the standard LS estimator.}

Following the definition of $t$ and $p$ in { LS93}, we define 
 \bear
        t_{12} & := & \farc{1}{\Nd}\left[\farc{\Gt_{12}}{\Gp_1\Gp_2}-1\right] \,,\nn\\
        \tur_{12} & := & \farc{1}{\Nr}\left[\farc{\Gt_{12}}{\Gp_1\Gp_2}-1\right] \ = \ \frac{\Nd}{\Nr}\,t_{12} \,,
 \nn\\
        p_{12} & := & \frac{2}{\Nd(\Nd-1)}\left[\frac{\delta_{12}}{(1+\xi_1)\Gp_1} - 2\farc{\Gt_{12}}{\Gp_1\Gp_2} + 1\right] \,,\nn\\
        \puc_{12} & := & \frac{1}{\Nd\Nr}\left[\frac{\delta_{12}}{\Gp_1} - 2\farc{\Gt_{12}}{\Gp_1\Gp_2} + 1\right] \,,\nn\\
        \pur_{12} & := & \frac{2}{\Nr(\Nr-1)}\left[\frac{\delta_{12}}{\Gp_1} - 2\farc{\Gt_{12}}{\Gp_1\Gp_2} + 1\right]  
   \,,\nn\\
        q_{12} & := & \farc{1}{\Nd}\frac{\Gt_{12}}{\Gp_1\Gp_2} \ = \ t_{12} + \farc{1}{\Nd} \,,\nn\\
        \qur_{12} & := & \farc{1}{\Nr}\frac{\Gt_{12}}{\Gp_1\Gp_2} \ = \ \tur_{12} + \farc{1}{\Nr} \,.
 \label{eq:tpq}
 \eear
For their diagonals ($\br_1=\br_2$), we write $t$, $\tr$, $p$, $\pc$, $\pr$, $q$, and $\qr$.
Thus, $t\equiv t_{11}\equiv t_{22}$, $\tr\equiv \tur_{11}\equiv \tur_{22}$ and so on.
(We use superscripts for the matrices, e.g., $\tur(\br_1,\br_2),$ and subscripts for their diagonals, e.g., $\tr(\br)$.)

{Using these definitions, \eqref{eq:alphabetaresults} becomes}
 \begin{gather}
        (1+\xi_1)(1+\xi_2)\langle\alpha_1\alpha_2\rangle \ \simeq \ (1+\xi_1)(1+\xi_2)(4t_{12} + p_{12}) \nn\\
          \qquad + 4\frac{\Nd-2}{(\Nd-1)}(\xi_{12}-\xi_1\xi_2)q_{12} \,,
    \nn\\
        \langle\beta_1\beta_2\rangle \ \simeq \ t_{12} + \tur_{12} + \puc_{12} + \farc{\Nd-1}{\Nd}\xi_{12}\qur_{12} \,,\nn\\   
        \langle\gamma_1\gamma_2\rangle \ = \ 4\tur_{12} + \pur_{12} \,,\nn\\
        \langle\alpha_1\beta_2\rangle \ \simeq \ 2t_{12} \,,\quad
        \langle\beta_1\gamma_2\rangle \ = \ 2\tur_{12} \,,\quad\mbox{and}\quad
        \langle\alpha_1\gamma_2\rangle \ = \ 0 \,. 
        \label{eq:abcvar}
 \end{gather}

{ The new part in $\langle\beta_1\beta_2\rangle$ due to finite size of the random catalog is
 \beq
        \Delta\langle\beta_1\beta_2\rangle \ \simeq \ \tur_{12} + \puc_{12} + \farc{\Nd-1}{\Nd}\xi_{12}\qur_{12} \,.
 \eeq
}
 
Thus only $\langle\alpha_1\alpha_2\rangle$ and $\langle\beta_1\beta_2\rangle$ are affected by $\xi(\br)$ (in our approximation its effect cancels in $\langle\alpha_1\beta_2\rangle$).  
The results for $\langle\gamma_1\gamma_2\rangle$, $\langle\beta_1\gamma_2\rangle$, and $\langle\alpha_1\gamma_2\rangle$ are exact.  The result for $\langle\alpha_1\alpha_2\rangle$ involves all three approximations mentioned above, $\langle\alpha_1\beta_2\rangle$ involves approximations (1) and (2), and $\langle\beta_1\beta_2\rangle$ involves approximation (1).

We refer to $p$, $\puc$, and $\pur$ as `Poisson' terms
and $t$ and $\tur$ as `edge' terms (the difference between $\Gt_{12}$ and $\Gp_1\Gp_2$ is due to edge effects).  While the Poisson terms are strongly diagonal dominated, the edge terms are not.
Since $\Nd t_{12} = \Nr \tur_{12} \ll 1$, the $q$ terms are much larger than the edge terms, but they get multiplied by $\xi_{12}-\xi_1\xi_2$ or $\xi_{12}$. In the limit $\Nr \rightarrow \infty$: $\langle\beta_1\gamma_2\rangle \rightarrow 0$, $\langle\gamma_1\gamma_2\rangle \rightarrow 0$, 
$\langle\beta_1\beta_2\rangle \rightarrow t_{12}$; $\langle\alpha_1\alpha_2\rangle,$ and also $\langle\alpha_1\beta_2\rangle$ are unaffected.

We see that $DD$-$DR$ and $DR$-$RR$ correlations arise from edge effects.  If we increase the density of data or random objects, the Poisson terms decrease as $N^{-2}$ but the edge terms decrease only as $N^{-1}$ so the edge effects are more important for a higher density of objects.

Doubling the bin size (combining neighboring bins) doubles $\Gp(\br)$ but makes $\Gt(\br_1,\br_2)$ four times as large, since triplets where one leg was in one of the original smaller bins and the other leg was in the other bin are now also included.  Thus, the ratio $\Gt_{12}/(\Gp_1\Gp_2)$ and $t$ are not affected, but the dominant term in $p$, $1/(1+\xi)\Gp$ is halved.  Edge effects are thus more important for larger bins.

\subsection{Results for the standard Landy--Szalay estimator}
\label{sec:LS}

Inserting the results for $\langle\alpha_1\alpha_2\rangle$ and so on~into Eqs. (\ref{eq:LSbias_abc}) and (\ref{eq:LScov_abc}),
we get that the expectation value of the standard LS estimator \eqref{eq:LS} is 
 \bear
        \langle\wxiLS\rangle = \xi + \left(\xi-1\right)(4\tr+\pr) + 4\tr \,.
 \label{eq:LSbias}
 \eear
This holds also for large $\xi$ and in the presence of three-point and four-point correlations.  A finite $R$ catalog thus introduces a bias $\left(\xi-1\right)(4\tr+\pr) + 4\tr = -\pr+(4\tr+\pr)\xi$; the edge ($\tr$) part of the bias cancels in the $\xi\rightarrow 0 $ limit.

For the covariance we get 
\beq
 \begin{split}
        &\mbox{Cov}\left[\wxiLS(\br_1),\wxiLS(\br_2)\right] \ \equiv \ \left\langle\wxiLS(\br_1)\wxiLS(\br_2)\right\rangle - \left\langle\wxiLS(\br_1)\right\rangle\left\langle\wxiLS(\br_2)\right\rangle \\
        &\quad \simeq \ (1+\xi_1)(1+\xi_2)p_{12} + 4\puc_{12} + (1-\xi_1)(1-\xi_2)\pur_{12} \\
        &\qquad + 4\xi_1\xi_2(t_{12}+\tur_{12}) \\
        &\qquad + 4\farc{\Nd-2}{\Nd-1}(\xi_{12}-\xi_1\xi_2)q_{12} + 4\farc{\Nd-1}{\Nd}\xi_{12}\qur_{12}  \,.    
 \end{split}
 \label{eq:LScov_standard}
 \eeq   
Because of the approximations made, this result for the covariance does not apply to the realistic cosmological case; not even for large separations $r$, where $\xi$ is small,
since large correlations at small $r$ increase the covariance also at large $r$.  However, this concerns only $\langle\alpha_1\alpha_2\rangle$ and $\langle\alpha_1\beta_2\rangle$.  Our focus here is on the additional covariance due to the size and handling of the random catalog, which
for standard LS is
 \beq
 \begin{split}
        &\Delta\mbox{Cov}\left[\wxiLS(\br_1),\wxiLS(\br_2)\right] \ \eqsim \
        4\Delta\langle\beta_1\beta_2\rangle + (1-\xi_1)(1-\xi_2)\langle\gamma_1\gamma_2\rangle \\
        &\qquad - 2(1-\xi_1)\langle\gamma_1\beta_2\rangle - 2(1-\xi_2)\langle\beta_1\gamma_2\rangle \\
        &\quad \simeq \ 4\puc_{12} + (1-\xi_1)(1-\xi_2)\pur_{12} +4\xi_1\xi_2\tur_{12} + 4\farc{\Nd-1}{\Nd}\xi_{12}\qur_{12} \,.
 \end{split}
 \label{eq:LSdeltacov_standard}
 \eeq
 
To zeroth order in $\xi$ the covariance is given by the Poisson terms and the edge terms cancel to first order in $\xi$.  This is the property for which the standard LS estimator was designed. To first order in $\xi$, the $q$ terms contribute.  This $q$ contribution involves the triplet correlation $\xi_{12}$, which, depending on the form of $\xi(\br)$, may be larger than $\xi_1$ or $\xi_2$.

If we try to save cost by using a diluted random catalog with $\Nrp \ll \Nr$ for $RR$ pairs, $\langle\gamma_1\gamma_2\rangle$ is replaced by $\langle\gamma'_1\gamma'_2\rangle = 4\turp_{12} + \purpp_{12}$ with $\Nrp$ in place of $\Nr$, but $\langle\beta_1\gamma_2'\rangle = \langle\beta_1\gamma_2\rangle$ and $\langle\beta_1\beta_2\rangle$ are unaffected, so that the edge terms involving randoms no longer cancel.  In Sect.~\ref{sec:mocks} we see that this is a large effect.  Therefore, one should not use dilution.

\section{Split random catalog}
\label{sec:split}

\subsection{Bias and covariance for the split method}

In the split method one has $\Ms$ independent smaller $R^\mu$ catalogs of size $\Nrp$ instead of one large random catalog $R$.  
Their union, $R,$ has a size of $\Nr = \Ms\Nrp$.  
The pair counts $DR(\br)$ and $RR'(\br)$ are calculated as
 \beq
        DR(\br)  :=  \sum_{\mu=1}^{\Ms}DR^\mu(\br) \quad\mbox{and}\quad
        RR'(\br)  :=  \sum_{\mu=1}^{\Ms}R^\mu R^\mu(\br) \,,
 \eeq
that is, pairs across different $R^\mu$ catalogs are not included in $RR'$.
The total number of pairs in $RR'$ is $\half \Ms\Nrp(\Nrp-1) = \half \Nr(\Nrp-1)$. Here,
$DR$ is equal to its value in standard LS.

The split Landy--Szalay estimator is
  \beq
        \wxi_\mathrm{split}(\br) \ := \ \farc{\Nr(\Nrp-1)}{\Nd(\Nd-1)}\frac{DD(\br)}{RR'(\br)} - \farc{\Nrp-1}{\Nd}\frac{DR(\br)}{RR'(\br)} + 1 \,.
 \eeq
Compared to standard LS,
$\langle\alpha_1\alpha_2\rangle$, $\langle\beta_1\beta_2\rangle$, and $\langle\alpha_1\beta_2\rangle$ 
are unaffected.
We construct
$\langle RR'\rangle$, 
$\langle RR'\cdot RR'\rangle$, and $\langle RR'\cdot DR\rangle$
from the standard LS results, bearing in mind that
the random catalog is a union of independent catalogs, arriving at
 \bear
        \langle\beta_1\gamma_2'\rangle \ &=& \  2\tur_{12}  \,,\nn \\
        \langle\gamma'_1\gamma'_2\rangle \ &=& \  4\tur_{12} + \purp_{12} \,,
 \eear
 where
  \beq
        \purp_{12} \ := \ \farc{\Nd(\Nd-1)}{\Nr(\Nrp-1)}\,p_{12} \ \equiv \ \farc{\Nr-1}{\Nrp-1}\,\pur_{12} \,.
 \eeq
The first is the same as in standard LS and dilution, but the second differs both from standard LS and from dilution, since it involves both $\Nr$ and $\Nrp$.
 
For the expectation value we get 
 \beq
        \langle\wxi_\mathrm{split}\rangle \ = \  \xi + \left(\xi-1\right)(4\tr+\prp) + 4\tr \,,
 \label{eq:split_bias}
 \eeq
so that the bias is $(\xi-1)(4\tr+\prp) + 4\tr \ = \ -\prp + (4\tr+\prp)\xi$.
In the limit $\xi\rightarrow0$ the edge part cancels, leaving only the Poisson term.

The covariance is 
 \beq
 \begin{split}
        &\mbox{Cov}\left[\wxi_\mathrm{split}(\br_1),\wxi_\mathrm{split}(\br_2)\right]  \\
        &\quad \simeq \ (1+\xi_1)(1+\xi_2)p_{12} + 4\puc_{12} + (1-\xi_1)(1-\xi_2)\purp_{12} \\      
        &\qquad + 4\xi_1\xi_2(t_{12}+\tur_{12})  \,.
 \end{split}
 \label{eq:cov_split}
 \eeq
The \emph{change} in the covariance compared to the standard LS method is
 \beq
 \begin{split}
        &\mbox{Cov}\left[\wxi^\mathrm{split}_1,\wxi^\mathrm{split}_2\right] - \mbox{Cov}\left[\wxi^\mathrm{LS}_1,\wxi^\mathrm{LS}_2\right] \\
        &\quad = \ (1-\xi_1)(1-\xi_2)(\purp_{12}-\pur_{12}) \,,
 \end{split}
 \label{eq:deltacov_split}
 \eeq
which again applies in the realistic cosmological situation.
Our main result is that in the split method the edge effects cancel and 
the bias and covariance are the same as for standard LS, 
except that the Poisson term $\pur$ from $RR$ is replaced with the larger $\purp$.

\subsection{Optimizing computational cost and variance of the split method}
\label{sec:optimize}

The bias is small compared to variance in our application { (see Fig.~\ref{fig:minerva_shell_loglog} for the theoretical result and Fig.~\ref{fig:minerva_bias} for an attempted bias measurement)}, and therefore we focus on variance as the figure of merit.
The computational cost should be roughly proportional to
 \beq
        \half \Nd^2\left(1+2\Mr+\frac{\Mr^2}{\Ms}\right) \ =: \ \half \Nd^2c \,,
 \label{eq:c-factor}
 \eeq
and the additional variance due to finite $R$ catalog in the $\xi\rightarrow 0$ limit becomes
 \beq
        \Delta\mbox{var} \approx \left(\frac{2}{\Mr}+\farc{\Ms}{\Mr^2}\right)p \ =: \ vp\,.
 \label{eq:v-factor}
 \eeq
Here, $\Nd$ and $p$ are fixed by the survey and the requested $\br$ binning, but we can vary $\Mr$ and $\Ms$ in the search for the optimal computational method.  In the above we defined the `cost' and `variance' factors $c$ and $v$.

We may ask two questions:
 \begin{enumerate}
        \item For a fixed level of variance $v$, which combination of $\Mr$ and $\Ms$ minimizes computational cost $c$?
        \item For a fixed computational cost $c$, which combination of $\Mr$ and $\Ms$ minimizes the variance $v$?
 \end{enumerate}
The answer to both questions is \citep{SlepianEisenstein}
 \beq
        \Ms = \Mr \qquad \Rightarrow\qquad c = 1+3\Mr \quad\mbox{and}\quad v = \farc{3}{\Mr} \,.
 \label{eq:optimal}
 \eeq 

Thus, the optimal version of the split method is the natural one where $\Nrp = \Nd$.  In this case the additional variance in the $\xi\rightarrow0$ limit becomes
 \beq
        \Delta\mbox{var}\ \approx \ \left(2\frac{\Nd}{\Nr} + \farc{\Nd}{\Nr}\right)p 
 ,\eeq
and the computational cost factor $\Nd^2+2\Nd\Nr+\Nr \Nrp$ becomes
 \beq
        \left(1 + 2\frac{\Nr}{\Nd} + \farc{\Nr}{\Nd}\right)\Nd^2 \,,
 \eeq
meaning that $DR$ pairs contribute twice as much as $RR$ pairs to the variance and also twice as much computational cost is invested in them.
The memory requirement for the random catalog is then the same as for the data catalog.
The cost saving estimate above is optimistic, since the computation involves some overhead not proportional to the number of pairs.

For small scales, where $\xi \gg 1$, the situation is different.  The greater density of $DD$ pairs due to the correlation requires a greater density of the $R$ catalog so that the additional variance from it is not greater. From Eq.~\eqref{eq:LSdeltacov_standard} we see that the balance of the $DR$ and the $RR$ contributions is different for large $\xi$  (the $\pc$ term vs.~the other terms).  We may consider recomputing $\wxi$ for the small scales using a smaller $r_\mathrm{max}$ and a larger $R$ catalog.  Considering just the Poisson terms ($\pc$ and $\pr$ or $\prp$) with a `representative' $\xi$ value, \eqref{eq:c-factor} and \eqref{eq:v-factor} become $c = 1+\xi+2\Mr+\Mr^2/\Ms$ and $v = 2/\Mr+(1-\xi)^2\Ms/\Mr^2$ which modifies the above result (Eq.~\ref{eq:optimal}) for  the optimal choice of $\Ms$ and $\Mr$ to 
 \beq
    \Ms = \farc{\Mr}{|\xi-1|} \,, \qquad\mbox{that is,}\qquad \Nrp \ = \ |\xi-1|\Nd \,.
 \eeq
This result is only indicative, since it assumes a constant $\xi$ for $r<r_\mathrm{max}$.  In particular, it does not apply for $\xi\approx 1$, because then the approximation of ignoring the $\qur$ and $\tur$ terms in \eqref{eq:LSdeltacov_standard} is not good.

\section{Tests on mock catalogs}
\label{sec:mocks}

\subsection{Minerva simulations and methodology}

The Minerva mocks are a set of 300 cosmological mocks produced with $N$-body simulations \citep{Grieb:2015bia, lippich19}, stored at five output redshifts $z\in\{2.0, 1.0, 0.57, 0.3, 0\}$.
The cosmology is flat $\Lambda$CDM with $\Omega_m = 0.285$, and we use the $z = 1$ outputs.  The mocks have $\Nd\approx 4\times10^6$ objects (``halos'' found by a friends-of-friend algorithm) in a box of $1500\hMpc$ cubed.  

To model the survey geometry of a redshift bin with $\Delta z \approx 0.1$ at $z \sim 1$, we placed the observer at comoving distance $2284.63\hMpc$ from the center of the cube and selected from the cube a shell $2201.34$--$2367.92\hMpc$ from the observer.
The comoving thickness of the shell is $166.58\hMpc$. 
The resulting mock sub-catalogs have $\Nd \approx 4.5\times10^5$ and
are representative of the galaxy number density
of the future Euclid spectroscopic galaxy catalog. 
 
We ignore peculiar velocities, that is, we perform our analysis in real space. Therefore, we consider results for the 1D
2PCF $\xi(r)$.
We estimated $\xi(r)$ up to $r_\mathrm{max} = 200\hMpc$ using $\Delta r = 1\hMpc$ bins.

We chose standard LS with $\Mr = 50$ as the reference method. 
In the following, LS without further qualification refers to this.  The random catalog was generated separately for each shell mock to measure their contribution to the variance.  For one of the random catalogs we calculated also triplets to obtain the edge effect quantity $\Nd t_{12} = \Gt_{12}/\Gp_1\Gp_2 - 1$.

While dilution can already be discarded on theoretical grounds, we show results {obtained using dilution}, since these results provide the scale for edge effects demonstrating the importance of eliminating them with a careful choice of method.
For the dilution and split methods we also used $\Mr = 50$ , and tried out dilution fractions $d := \Nrp/\Nr = 0.5, 0.25, 0.14$ and split factors $\Ms = 4, 16, 50$ (chosen to have similar  pairwise computational costs).  In addition, we considered  standard LS with $\Mr=25$, which has the same number of $RR$ pairs as $d = 0.5$ and $\Ms = 4$, but only half the number of $DR$ pairs; and standard LS with $\Mr=1$ to demonstrate the effect of a small $\Nr$.


The code used to estimate the 2PCF implements a highly optimized pair-counting method, specifically designed for the search of object pairs in a given range of separations.{ In particular, the code provides two alternative counting
methods, the {\em chain-mesh} and the {\em kd-tree}.  Both methods measure the exact number of object pairs in separation bins, without any approximation. However, since they implement different algorithms to search for pairs, they perform differently at different scales, both in terms of CPU time and memory usage. Overall, the efficiency of the two methods depends on the ratio between the scale range of the searching region and the maximum separation between the objects in the catalog.}

{ The {\em kd-tree} method first constructs a space-partitioning data structure that is filled with catalog objects. The minimum and maximum separations probed by the objects are kept in the data structure and are used to prune object pairs with separations outside the range of interest. The tree pair search is performed through the dual-tree method in which cross-pairs between two dual trees are explored. This is an improvement in terms of exploration time over the single-tree method.}

{ On the other hand, in the {\em chain-mesh} method the catalog is divided in cubic cells of equal size, and the indexes of the objects in each cell are stored in vectors. To avoid counting object pairs with separations outside the interest range, the counting is performed only on the cells in a selected range of distances from each object. The {\em chain-mesh} algorithm has been} imported from the CosmoBolognaLib, a large set of {\em free software} C++/python libraries for cosmological calculations \citep{2016AC....14...35M}.

For our test runs we used the {\em chain-mesh} method.

\subsection{Variance and bias}
\label{sec:var_bias}

\begin{figure}
  \resizebox{\hsize}{!}{\includegraphics{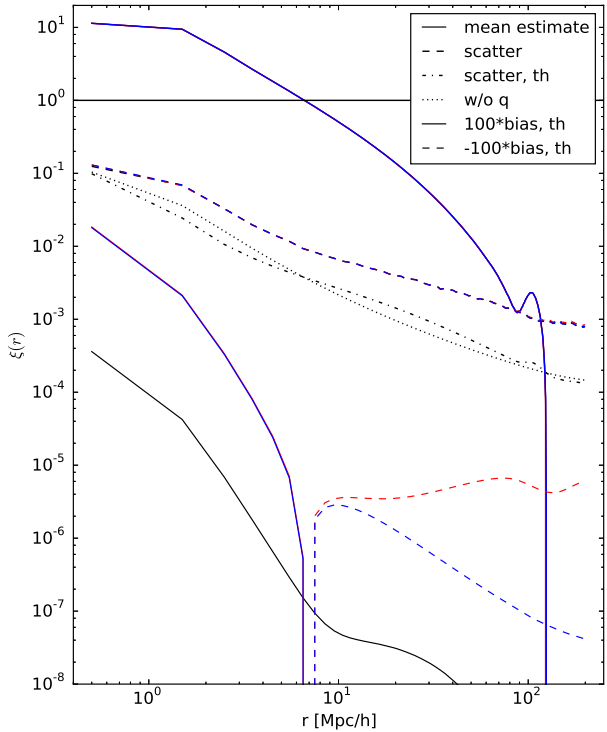}}
  \caption{Mean $\xi(r)$ estimate and the scatter \hx{and theoretical bias} of the estimates for different estimators.   The dash-dotted line, our theoretical result for the scatter of the LS method, underestimates the scatter, since higher-order correlations in the $D$ catalog are ignored.    The dotted line is without the contribution of the $q$ terms, and is dominated by the Poisson ($p$) terms.  
The bias is multiplied by $100$ so the curves can be displayed in a more compact plot.
For the measured mean and scatter, and the theoretical bias we plot standard LS in black, dilution with $d = 0.14$ in red, and split with $M_s=50$ in blue.  For the mean and scatter the difference between the methods is not visible in this plot.  The differences in the mean estimate are shown in Fig.~\ref{fig:minerva_bias}.  The differences in scatter (or its square, the variance) are shown in Fig~\ref{fig:minerva_vardif_lin2}.  For the theoretical bias the difference between split and dilution is not visible at small $r$ ($\xi(r) >1$), where the bias is positive.}
  \label{fig:minerva_shell_loglog}
\end{figure}

In Fig.~\ref{fig:minerva_shell_loglog} we show the mean (over the 300 mock shells) estimated correlation function and the scatter (square root of the variance) of the estimates using the LS, split, and dilution methods; our theoretical approximate result for the scatter for LS; \hx{and our theoretical result for bias for the different methods}.

The theoretical result for the scatter is shown with and without the $q$ terms, which include the triplet correlation $\xi_{12}$, for which we used here the approximation $\xi_{12} \approx \xi(\max(r_1,r_2))$.
This 
behaves as expected, that is, it underestimates the variance, since we neglected the higher-order correlations in the $D$ catalog.  Nevertheless, it (see the dash-dotted line in Fig.~\ref{fig:minerva_shell_loglog}) has similar features to the measured variance (dashed lines).

\begin{figure}
  \resizebox{\hsize}{!}{\includegraphics{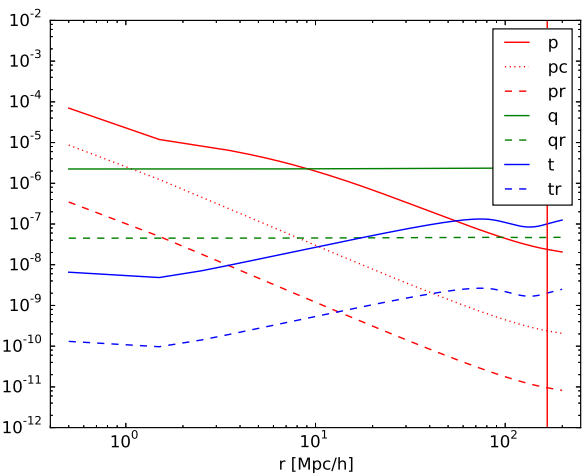}}
  \caption{ Quantities $p$, $p_c$, $p_r$, $q$, $q_r$, $t$, and $t_r$ for the Minerva shell.  The values for the first bin are noisy.  The vertical red line marks $r=L$.}
  \label{fig:minerva_shell_pqt}
\end{figure}

{ In Fig.~\ref{fig:minerva_shell_pqt} we plot the diagonals of the $p$, $t$, and $q$ quantities. This shows how their relative importance changes with separation scale.  It also confirms that our initial assumption on small relative fluctuations is valid in this simulation case.}

\begin{figure}
  \resizebox{\hsize}{!}{\includegraphics{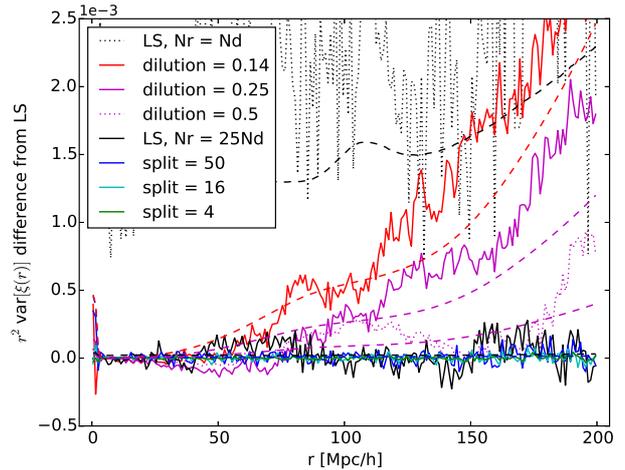}}
  \caption{Measured difference from LS of the variance of different estimators, multiplied by $r^2$.  Dashed lines are our theoretical results.}
  \label{fig:minerva_vardif_lin2}
\end{figure}

Consider now the variance differences (from standard LS with $\Mr = 50$), for which our theoretical results should be accurate. 
Figure~\ref{fig:minerva_vardif_lin2}
compares the measured variance difference to the theoretical result.  For the diluted estimators and LS with $\Mr = 1$ the measured result agrees with theory, although clearly the measurement with just 300 mocks is rather noisy.  For the split estimators and LS with $\Mr = 25$ the difference is too small to be appreciated with 300 mocks, but at least the measurement does not disagree with the theoretical result. 
\begin{figure}
  \resizebox{\hsize}{!}{\includegraphics{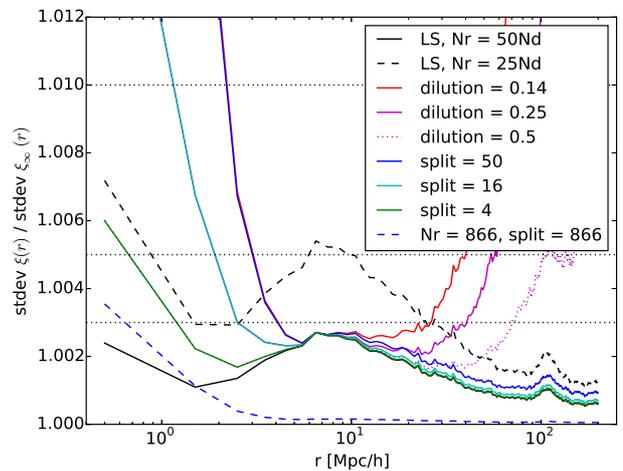}}
  \caption{Theoretical estimate of the scatter of the $\xi$ estimates divided by the scatter in the $\Nr\rightarrow\infty$ limit. The dotted lines correspond to 0.3\%, 0.5\%, and 1\% increase in scatter.  For $r<10\hMpc$ there is hardly any difference between split and dilution, the curves lie on top of each other; whereas for larger $r$ split is much better.}
  \label{fig:minerva_scatter_ratio_to_infty_th_log_zoom}
\end{figure}

In Fig.~\ref{fig:minerva_scatter_ratio_to_infty_th_log_zoom} we show the relative theoretical increase in scatter compared to the best possible case, which is LS in the limit $\Mr\rightarrow\infty$.  
Since we do not have a valid theoretical result for the total scatter, we estimate it by subtracting the theoretical difference from LS with $ \Mr = 50$ from the measured variance of the latter.  

At scales $r\lesssim 10\hMpc$ the theoretical prediction is about the same for dilution and split and neither method looks promising for $r\ll 10\hMpc$ where $\xi\gg1$.  This suggests that for optimizing cost and accuracy, a different method should be used for smaller scales than that used for large scales.  The number of $RR$ pairs with small separations is much less. Therefore, for the small-scale computation there is no need to restrict the computation to a subset of $RR$ pairs, or alternatively, one can afford to increase $\Mr$.
For the small scales, we may consider the split method with increased $\Mr$ as an alternative to standard LS.  We have the same number of pairs to compute as in the reference LS case, if we use $\Mr = 866$ and $\Ms = 866$.  We added this case to Fig.~\ref{fig:minerva_scatter_ratio_to_infty_th_log_zoom}.  It seems to perform better than LS at intermediate scales, but for the smallest scales LS has the smaller variance.  This is in line with our conclusion in Sect.~\ref{sec:optimize}, which is that when $\xi\gg1$, it is not optimal to split the $R$ catalog into small subsets.

\begin{figure}
  \resizebox{\hsize}{!}{\includegraphics{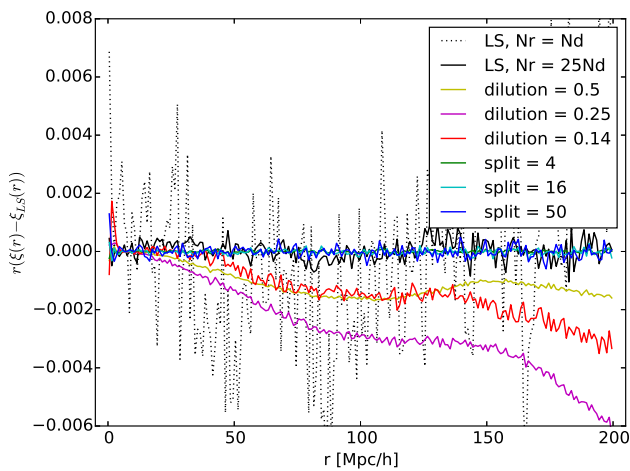}}
  \caption{\hx{Differences between the mean $\xi(r)$ estimate and that from the LS,
multiplied by $r$ to better display all scales.  This measured difference is not the true bias, which is too small to measure with 300 mocks, and is mainly due to random error of the mean. The results for dilution appear to reveal a systematic bias, but this is just due to strong error correlations between nearby bins;
for different subsets of the 300 mocks the mean difference is completely different.}}
  \label{fig:minerva_bias}
\end{figure}

We also compared the differences in the mean estimate from the different estimators to our theoretical results on the bias differences (see Fig.~\ref{fig:minerva_bias}), but the theoretical bias differences are much smaller than the expected error of the mean from 300 mocks; and we simply confirm that the differences we see are consistent with the error of the mean and thus consistent with the true bias being much smaller.
We  also performed tests with completely random ($\xi = 0$) mocks, and with a large number ($10\,000$) of mocks confirmed the theoretical bias result for the different estimators in this case.  Since the bias is too small to be interesting we do not report these results in more detail here.

However, we note that for the estimation of the 2D 2PCF and its multipoles, the 2D bins will contain a smaller number of objects than the 1D bins of these test runs and therefore the bias is larger.  Using the theoretical results \eqref{eq:LSbias} or \eqref{eq:split_bias} the bias can be removed afterwards with accuracy depending on how well we know the true $\xi$.

\subsection{Computation time and variance}

\begin{figure}
    \centering
    \resizebox{\hsize}{!}{\includegraphics{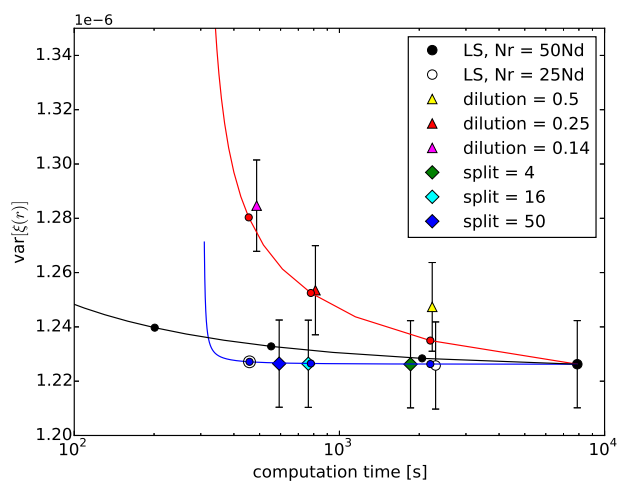}}
    \caption{Measured variance (mean variance over the range $r = 80$--$120\hMpc$) vs. computational cost (mean computation time) for the different methods (markers with error bars) and our theoretical prediction (solid lines).  
The solid lines (blue for the split method, red for dilution, and black for standard LS with $\Mr\leq50$) are our theoretical predictions 
for the increase in variance and computation time ratio when compared to the standard LS, $\Mr=50$, case, and the dots on the curves correspond to the measured cases (except for LS they are, from right to left, $\Mr = 25$, $12.5$, and $(50/7)$; only the first of which was measured). The curve for split ends at $M_s=2500$; the optimal case, $\Ms=\Mr$, is the circled dot. The error bars for the variance measurement are naive estimates that do not account for error correlations between bins.  The theoretical predictions overestimate the cost savings (data points are to the right of the dots on curves; except for the smaller split factors, where the additional speed-up compared to theory is related to some other performance differences between our split and standard LS implementations).  This plot would have a different appearance for other $r$ ranges.
}
    \label{fig:minerva_var_vs_cost}
\end{figure}

\hx{The test runs were made using a single full 24-core node for each run.}
Table~\ref{tab:time_and_var} shows the mean computation time and mean estimator variance for different $r$ ranges for the different cases we tested.  Of these $r$ ranges, the $r = 80$--$120\hMpc$ is perhaps the most interesting, since it contains the important BAO scale.  Therefore, we plot the mean variance at this range versus mean computation time in Fig.~\ref{fig:minerva_var_vs_cost}, together with our theoretical predictions.  The theoretical estimate for the computation time for other dilution fractions and split factors is
 \beq
    (1+24.75d^2)\ 306\,\mbox{s} \qquad\mbox{and}\qquad (1+24.75/\Ms^2)\ 306\,\mbox{s}
 ,\eeq
assuming $\Mr = 50$.
For standard LS with other random catalog sizes, the computation time estimate is
 \beq
    (1+2\Mr+\Mr^2)\ 3.03\,\mbox{s} \,.
 \eeq

\begin{table}
\centering
    \caption{Mean computation time over the 300 runs and the mean variance over four different ranges of $r$ bins (given in units of $\hMpc$) for each method.  The first three are standard LS. The variance cannot be measured accurately enough from 300 realizations to correctly show all the differences between methods.  Thus the table shows some apparent improvements (going from $\Mr = 50$ to $\Mr = 25$ and from $\Ms = 4$ to $\Ms = 16$), which are not to be taken as real.  See Fig.~\ref{fig:minerva_var_vs_cost} for the fifth vs. second column with error bars.}
    \begin{tabular}{lrcccc}
    \hline\hline
    method & time & \multicolumn{4}{c}{mean variance} \\
    & [s] & $0$--$2$ & $5$--$15$ & $80$--$120$ & $150$--$200$ \\
    &&  [$\times10^{-2}$] & [$\times10^{-5}$] & [$\times10^{-6}$] & [$\times10^{-7}$] \\ 
    \hline
    $\Mr = 50$ & 7889 & 1.01 & 4.42 & 1.23 & 7.01 \\
    $\Mr = 25$ & 2306 & 1.01 & 4.39 & 1.23 & 7.04 \\
    $\Mr = 1 $  & 16.5 & 5.96 & 5.53 & 1.46 & 8.05 \\
    $d = 0.5$  & 2239 & 1.01 & 4.41 & 1.25 & 7.11 \\
    $d = 0.25$ & 812  & 1.02 & 4.41 & 1.25 & 7.42 \\
    $d = 0.14$ & 487  & 1.08 & 4.41 & 1.28 & 7.72 \\
    $\Ms = 4$ & 1854 & 1.01 & 4.42 & 1.23 & 7.02 \\
    $\Ms = 16$ & 763  & 1.00 & 4.42 & 1.23 & 7.01 \\
    $\Ms = 50$ & 593  & 1.09 & 4.42 & 1.23 & 7.02 \\
    \hline\hline
    \end{tabular}
    \label{tab:time_and_var}
\end{table}

\section{Conclusions}
\label{sec:conclusions}

The computational time of the standard Landy--Szalay estimator is dominated by the $RR$ pairs. However, except at small scales where correlations are large, these make a negligible contribution to the expected error compared to the contribution from the $DD$ and $DR$ pairs.  Therefore, a substantial saving of computation time with an insignificant loss of accuracy can be achieved by counting a smaller subset of $RR$ pairs.  

We considered two ways to reduce the number of $RR$ pairs: dilution and split.  In dilution, only a subset of the $R$ catalog is used for $RR$ pairs.  In split, the $R$ catalog is split into a number of smaller subcatalogs, and only pairs within each subcatalog are counted.  We derived theoretical results for the additional estimator covariance and bias due to the finite size of the random catalog for these different variants of the LS estimator, extending in many ways the original results by \citet{Landy:1993yu}, who worked in the limit of an infinite random catalog.  We tested our results using 300 mock data catalogs, representative of the $z = 0.95$--$1.05$ redshift range of the future Euclid survey.  The split method maintains the property the Landy--Szalay estimator was designed for, namely cancelation of edge effects in bias and variance (for $\xi = 0$), whereas dilution loses this cancellation and therefore should not be used.

For small scales, where correlations are large, 
one should not reduce $RR$ counts as much.  The natural dividing line is the scale $r$ where $\xi(r) = 1$.  Interestingly, the difference in bias and covariance between the different estimators (split, dilution, and LS) vanishes when $\xi = 1$.  We recommend the natural version of the split method, $\Ms = \Mr$, for large scales where $|\xi| < 1$.   This leads to a saving in computation time by more than a factor of ten (assuming $\Mr = 50$) with a negligible effect on variance and bias.  For small scales, where $\xi > 1$, one should consider using a larger random catalog and one can use either the standard LS method or the split method with a more modest split factor.  Because the number of pairs with these small separations is much smaller, the computation time is not a similar issue as for large separations.

\eb{The results of our analysis will have an impact also on the computationally more demanding task of covariance matrix estimation. However, since in that case the exact computational cost is determined by 
the balance of data catalogs and random catalogs, which does not need to be the same as for the individual two-point correlation estimate, we postpone a quantitative analysis to a future, dedicated study.}
The same kind of methods can be applied to higher-order statistics (three-point and four-point correlation functions) to speed up their estimation \citep{SlepianEisenstein}.

\begin{acknowledgements} 
We thank Will Percival and Cristiano Porciani for useful discussions.  The 2PCF computations were done at the Euclid Science Data Center Finland (SDC-FI, urn:nbn:fi:research-infras-2016072529), for whose computational resources we thank CSC -- IT Center for Science, the Finnish Grid and Cloud Computing Infrastructure (FGCI, urn:nbn:fi:research-infras-2016072533), and the Academy of Finland grant 292882.  This work was supported by the Academy of Finland grant 295113.  VL was supported by the Jenny and Antti Wihuri Foundation, AV by the V\"{a}is\"{a}l\"{a} Foundation, AS by the Magnus Ehrnrooth Foundation, and JV by the Finnish Cultural Foundation.  We also acknowledge travel support from the Jenny and Antti Wihuri Foundation. VA acknowledges funding from the European Union's Horizon 2020 research and innovation programme under grant agreement No 749348. FM acknowledges the grants ASI n.I/023/12/0 ``Attivit`a relative alla fase B2/C per la missione Euclid'' and PRIN MIUR 2015
``Cosmology and Fundamental Physics: illuminating the Dark Universe with Euclid''.
\end{acknowledgements}

\bibliographystyle{aa} 

\begin{thebibliography}{99}
\bibitem[\protect\citeauthoryear{Alam et al.}{2017}]{alam17} Alam S. et al. 2017, MNRAS {470}, 2617
\bibitem[Alcock~\&~Paczy\'{n}ski(1979)]{ap79} Alcock, C. \& Paczy\'{n}ski, B. 1979, Nature {281}, 358
\bibitem[Alonso(2012)]{2012arXiv1210.1833A} Alonso, D.\ 2012, arXiv:1210.1833
\bibitem[Ata et al.(2018)]{ata18} Ata, M. et al. 2018, MNRAS {473}, 4773
\bibitem[Anderson et al.(2012)]{anderson12} Anderson, L. et al. 2012, MNRAS {427}, 3435
\bibitem[Anderson et al.(2014)]{anderson14} Anderson, L. et al. 2014, MNRAS {441}, 24
\bibitem[Bautista et al.(2018)]{bautista18} Bautista, J.E. et al. 2018, ApJ {863}, 110
\bibitem[Bernstein(1994)]{bernstein94} Bernstein, G.M. 1994, ApJ {424}, 569
\bibitem[Beutler et al.(2011)]{beutler11} Beutler, F. et al. 2011, MNRAS {416}, 3017
\bibitem[Beutler et al.(2012)]{beutler12} Beutler, F. et al. 2012, MNRAS {423}, 3430
\bibitem[Blake et al.(2011)]{blake11} Blake, C. et al. 2011, MNRAS {418}, 1707
\bibitem[Cole et al.(2005)]{cole05} Cole, S. et al. 2005, MNRAS {362}, 505
\bibitem[Davis~\&~Peebles(1983)]{dp83} Davis, M. \& Peebles, P.J.E 1983, ApJ {267}, 465
\bibitem[Demina et al.(2018)]{Demina2018} Demina, R., Cheong, S., BenZvi, S., \& Hindrichs, O. 2018, MNRAS {480}, 49
\bibitem[Eisenstein et al.(2005)]{eisenstein05} Eisenstein, D.J. et al. 2005, ApJ {633}, 560
\bibitem[Grieb et al.(2016)]{Grieb:2015bia} Grieb, J.N., S\'{a}nchez, A.G, Salvador-Albornoz, S., \&  Dalla Vecchia, C. 2016, MNRAS {457}, 1577
\bibitem[Guzzo et al.(2008)]{guzzo08} Guzzo, L. et al. 2008, Nature {451}, 541
\bibitem[Hamilton(1993)]{hamilton93} Hamilton, A.J.S. 1993, ApJ {417}, 19
\bibitem[Hewett(1982)]{hewett82}Hewett, H.C. 1982, MNRAS {201}, 867
\bibitem[Hou et al.(2018)]{hou18} Hou, J. et al. 2018, MNRAS {480}, 2521
\bibitem[Jarvis(2015)]{2015ascl.soft08007J} Jarvis, M.\ 2015, Astrophysics Source Code Library, ascl:1508.007
\bibitem[Kaiser(1987)]{kaiser87} Kaiser, N. 1987, MNRAS {227}, 1
\bibitem[Kerscher(1999)]{Kerscher99} Kerscher, M. 1999, A\&A {343}, 333
\bibitem[Kerscher(2000)]{Kerscher00} Kerscher, M., Szapudi, I., \& Szalay, A.S. 2000, ApJ {535}, L13
\bibitem[Landy~\&~Szalay(1993)]{Landy:1993yu} Landy, S.D. \& Szalay, A.S. 1993, ApJ {412}, 64 
\bibitem[\protect\citeauthoryear{Laureijs et al.}{2011}]{laureijs11} Laureijs R. et al. 2011, arXiv:1110.3193
\bibitem[Lippich et al.(2019)]{lippich19} Lippich, M. et al. 2019, MNRAS {482}, 1786
\bibitem[Marulli et al.(2016)]{2016AC....14...35M} Marulli, F., Veropalumbo, A., \& Moresco, M.\ 2016, Astronomy and Computing {14}, 35 
\bibitem[Moore et al.(2000)]{moore01} Moore, A. et al. 2000, arXiv:astro-ph/0012333
\bibitem[Peacock et al.(2001)]{peacock01} Peacock, J.A. et al. 2001, Nature {410}, 169
\bibitem[Peebles~\&~Hauser(1974)]{ph74} Peebles, P.J.E. \& Hauser, M.G. 1974, ApJS {28}, 19
\bibitem[Percival et al.(2010)]{percival10} Percival, W.J. et al. 2010, MNRAS {401}, 2148
\bibitem[Pezzotta et al.(2017)]{pezzotta17} Pezzotta, A. et al. 2017, A\&A {604}, A33
\bibitem[Reid et al.(2012)]{reid12} Reid, B.A. et al. 2012, MNRAS {426}, 2719
\bibitem[Ross et al.(2015)]{ross15} Ross, A.J. et al. 2015, MNRAS {449}, 835
\bibitem[Ross et al.(2017)]{ross17} Ross, A.J. et al. 2017, MNRAS {464}, 1168
\bibitem[Ruggeri et al.(2019)]{ruggeri19} Ruggeri, R. et al. 2019, MNRAS {483}, 3878
\bibitem[Slepian~\&~Eisenstein(2015)]{SlepianEisenstein} Slepian, Z. \& Eisenstein, D.J. 2015, MNRAS {454}, 4142
\bibitem[de la Torre et al.(2017)]{delatorre17} de la Torre, S. et al. 2017, A\&A {608}, A44
\bibitem[Vargas-Maga\~{n}a et al.(2018)]{vargas18} Vargas-Maga\~{n}a, M. et al. 2018, MNRAS {477}, 1153
\bibitem[Wall~\&~Jenkins(2012)]{WallJenkins} Wall, J.V. \& Jenkins, C.R. 2012, Practical Statistics for Astronomers (Cambridge University Press), Sec.~10.4.1
\bibitem[Zarrouk et al.(2018)]{zarrouk18} Zarrouk, P. et al. 2018, MNRAS {477}, 1639
\bibitem[Zehavi et al.(2011)]{zehavi18} Zehavi, I. et al. 2011, ApJ {736}, 59

\end{thebibliography}

\appendix
\section{Derivation examples}
\label{sec:examples}


As examples of how  the  variances of the different deviations, $\langle\alpha\beta\rangle$ and so on,~are derived, following the method presented in { LS93}, we give three of the cases here: $\langle\alpha_1\alpha_2\rangle$ (common for all the variants of LS), and $\langle\beta_1\gamma'_2\rangle$ and $\langle\gamma'_1\gamma'_2\rangle$ for the split method.   The rest are calculated in a similar manner.  

To derive the $\langle DD_1\ DD_2 \rangle$ appearing in the $\langle\alpha_1\alpha_2\rangle$ of \eqref{eq:abcvar} we start from
 \beq
        \langle DD_1\ DD_2\rangle \ = \ \sum_{i<j}^K \sum_{k<l}^K\langle n_in_jn_kn_l\rangle \Theta^{ij}(\br_1)\Theta^{kl}(\br_2) \,,
 \label{eq:DDDDsum}
 \eeq 
where both $i,j,$ and $k,l$ sum over all microcell pairs; $n_i = 1\mbox{\ or\ }0$ is the number of galaxies in microcell $i$.
There are three different cases for the terms $\langle n_in_jn_kn_l\rangle$ depending on how
many indices are equal ($i\neq j$ and $k\neq l$ for all of them).

The first case (quadruplets, i.e., two pairs of microcells) is when $i,j,k,l$ are all different.  We use  $\sum^*_{ijkl}$  to denote this part of the sum.  There are $\half K(K-1)\times\half(K-2)(K-3) = \quarter K^4$ (we work in the limit $K\rightarrow\infty$) such terms and they have 
 \bear
 & &    \hspace{-2em}\langle n_in_jn_kn_l\rangle \nn\\
        & = &  \farc{\Nd}{K}\farc{\Nd\! -\! 1}{K\! -\! 1}\farc{\Nd\!-\! 2}{K\!-\! 2}\farc{\Nd\!-\! 3}{K\!-\! 3}\Bigl\langle(1+\delta_i)(1+\delta_j)(1+\delta_k)(1+\delta_l)\Bigr\rangle
        \nn\\
        & = &  \frac{\Nd(\Nd\! -\! 1)(\Nd\! -\! 2)(\Nd\! -\! 3)}{K^4}\Bigl[1+\langle\delta_i\delta_j\rangle+\langle\delta_k\delta_l\rangle+\langle\delta_i\delta_k\rangle+ \nn\\
        & & +\langle\delta_i\delta_l\rangle+\langle\delta_j\delta_k\rangle+\langle\delta_j\delta_l\rangle+ \nn\\
        & & +
        \langle\delta_i\delta_j\delta_k\rangle+\langle\delta_i\delta_j\delta_l\rangle+\langle\delta_i\delta_k\delta_l\rangle+\langle\delta_j\delta_k\delta_l\rangle+\langle\delta_i\delta_j\delta_k\delta_l\rangle\Bigr] \nn\\
        & = & \frac{\Nd(\Nd\! -\! 1)(\Nd\! -\! 2)(\Nd\! -3\! )}{K^4}\Bigl[1+\xi(\br_{ij})+\xi(\br_{kl})+\xi(\br_{ik})+ \nn\\
        & & + \xi(\br_{il})+\xi(\br_{jk})+\xi(\br_{jl})+ \nn\\
  & & + \zeta(\bx_i,\bx_j,\bx_k)+\zeta(\bx_i,\bx_k,\bx_l)+\zeta(\bx_i,\bx_j,\bx_l)+\zeta(\bx_j,\bx_k,\bx_l)+    \nn\\   
        & & + \xi(\br_{ij})\xi(\br_{kl})+\xi(\br_{ik})\xi(\br_{jl})+\xi(\br_{il})\xi(\br_{jk}) \nn\\
        & & +\eta(\bx_i,\bx_j,\bx_k,\bx_l)\Bigr] \,,
 \label{eq:quadruplets}
 \eear
where $\delta_i := \delta(\bx_i$) is the density perturbation (relative to the actual mean density of galaxies in the survey),  $\zeta$ is the three-point correlation, and $\eta$ is the connected (i.e., the part that does not arise from the two-point correlation) four-point correlation.  
The fraction of microcell quadruplets (pairs of pairs) that satisfy $\br_{ij} \equiv \bx_j-\bx_i \in \br_1$ and $\br_{kl} \in \br_2$ is $\Gp(\br_1)\Gp(\br_2) =: \Gp_1\Gp_2$. In the limit of large $K$ the number of other index quadruplets is negligible compared to those where all indices have different values, so we have
 \beq
        \sum^\ast_{ijkl} \Theta^{ij}_1\Theta^{kl}_2 \ = \ \farc{K(K-1)(K-2)(K-3)}{4}\Gp_1\Gp_2 \ = \ \farc{K^4}{4}\Gp_1\Gp_2\,.
 \label{eq:four_index}
 \eeq
For the connected pairs $(i,j)$ and $(k,l)$ we have $\xi(\br_{ij}) = \xi(\br_1) \equiv \xi_1$ and $\xi(\br_{kl}) = \xi(\br_2) \equiv \xi_2$.

The second case (triplets of microcells) is when $k$ or $l$ is equal to $i$ or $j$.  We denote this part of the sum by 
 \beq
        \sum^*_{ijk}\langle n_in_jn_k\rangle \Theta_1^{ik}\Theta_2^{jk} \,.
 \eeq
It turns out that it goes over all ordered combinations of $\{i,j,k\}$, where $i,j,k$ are all different, exactly once, so there are $K(K-1)(K-2)) = K^3$ such terms (triplets).  Here
 \bear
         & &    \hspace{-2em}\langle n_in_jn_k\rangle \nn\\
         & = & \farc{\Nd}{K}\farc{\Nd\! -\! 1}{K\! -\! 1}\farc{\Nd\! -\! 2}{K\! -\! 2}\Bigl\langle(1+\delta_i)(1+\delta_j)(1+\delta_k)\Bigr\rangle
        \nn\\
        & = & \frac{\Nd(\Nd\! -\! 1)(\Nd\! -\! 2)}{K^3}\Bigl[1 +\langle\delta_i\delta_k\rangle+\langle\delta_j\delta_k\rangle+\langle\delta_i\delta_j\rangle + \langle\delta_i\delta_j\delta_k\rangle\Bigr] \nn\\
        & = & \frac{\Nd(\Nd\! -\! 1)(\Nd\! -\! 2)}{K^3}\Bigl[1+\xi(\br_{ik})+\xi(\br_{jk})+\xi(\br_{ij})+ \nn\\
        & & +\zeta(\bx_i,\bx_j,\bx_k)\Bigr] 
 \label{eq:triplets}
 ,\eear
and
 \beq
   \sum^\ast_{ijk}\Theta^{ik}_1\Theta^{jk}_2 = K^3\Gt_{12} \,.
 \eeq
 
The third case (pairs of microcells) is when $i=k$ and $j=l$.  This part of the sum becomes
 \beq
        \sum_{i<j} \langle n_in_j\rangle \Theta^{ij}_1\Theta^{ij}_2 \,,
 \eeq
where
 \beq
        \langle n_in_j\rangle \ = \ \farc{\Nd(\Nd-1)}{K^2}\left[1+\xi(\br_{ij})\right]
 ,\eeq
and
 \beq
        \sum_{i<j}\Theta^{ij}_1\Theta^{ij}_2 \ = \ \delta_{12}\farc{K^2}{2}\Gp_1 \,,
 \eeq
that is, the sum vanishes unless the two bins are the same, $\br_1 = \br_2$.

We now apply the three approximations listed in Sect.~\ref{sec:qta}:
1) $\xi(\br_{ik}) =  \xi(\br_{il}) = \xi(\br_{jk}) = \xi(\br_{jl}) = 0$ in \eqref{eq:quadruplets};
2) $\zeta = 0$ in \eqref{eq:quadruplets} and \eqref{eq:triplets};
3) $\eta = 0$ in \eqref{eq:quadruplets}.
We obtain
 \bear
        & &     \hspace{-2em}\langle DD_1\ DD_2 \rangle \nn\\
        & \simeq & \quarter \Nd(\Nd-1)(\Nd-2)(\Nd-3)(1+\xi_1)(1+\xi_2)\Gp_1\Gp_2 +  \nn\\        
        & & + \Nd(\Nd-1)(\Nd-2)(1+\xi_1+\xi_2+\xi_{12})\Gt_{12} + \nn\\
        & & + \half \Nd(\Nd-1)\delta_{12}(1+\xi_1)\Gp_1 
 ,\eear
and using $\langle DD\rangle$ from \eqref{eq:pair_expect} we arrive at the $\langle\alpha_1\alpha_2\rangle$ result given in Eq.~\eqref{eq:abcvar}.

For the split method, we give the calculation for 
 \bear
        \langle\gamma'_1\beta'_2\rangle & = & \farc{\langle RR'_1 \ DR_2 \rangle - \langle RR'_1  \rangle\langle DR_2 \rangle}{\langle RR'_1  \rangle\langle DR_2 \rangle} \,,\nn\\
        \langle\gamma'_1\gamma'_2\rangle & = &  \farc{\langle RR'_1\ RR'_2 \rangle - \langle RR'_1\rangle\langle RR'_2\rangle}{\langle RR'_1\rangle\langle RR'_2\rangle}
 \eear
below.

First the $\langle\gamma'_1\beta_2\rangle$:
 \beq
        \langle RR'_1 \ DR_2 \rangle \ = \ \sum_{\mu=1}^{\Ms}\langle R^\mu R^\mu_1\ DR_2\rangle \ = \ \Ms\langle  R^\mu R^\mu_1\ DR_2\rangle \,,
 \eeq
where
 \beq
        \langle R^\mu R^\mu_1\ DR_2\rangle \ = \ \sum_{i<j}\sum_{k\neq l}\langle s_is_jn_kr_l\rangle \Theta^{ij}_1\Theta^{kl}_2 \,,
 \eeq
$s_i$ is the number (0 or 1) of $R^\mu$ objects in microcell $i$, and $r_l$ is the number of $R$ objects in microcell $l$.

There are $\half K^4$ quadruplet terms, for which (see \eqref{eq:four_index}) $\sum^\ast\Theta\Theta = \half K^4\Gp_1\Gp_2$, and
 \beq
        \langle s_is_jn_kr_l\rangle \ = \ \frac{\Nrp(\Nrp-1)\Nd(\Nr-2)}{K^4} \,.
 \eeq
 
Triplets where $i$ or $j$ is equal to $k$ have $\langle s_is_kn_kr_l\rangle = 0$, since $s_kn_k = 0$ always (we cannot have two objects, from $R^\mu$ and $D$, in the same cell).
There are $K^3$ triplet terms where $i$ or $j$ is equal to $l$: for them $\sum^\ast\Theta\Theta = K^3\Gt_{12}$, and
 \beq
        \langle s_is_ln_kr_l\rangle \ = \ \langle s_is_ln_k\rangle \ = \ \frac{\Nrp(\Nrp-1)\Nd}{K^3}
 ,\eeq
where if $s_l =1$ then also $r_l=1$ since $R^\mu \subset R$.

Pairs where $(i,j) = (k,l)$ or $(i,j) = (l,k)$ have $\langle s_ks_ln_kr_l\rangle = 0$, since again we cannot have two different objects in the same cell.
Thus 
 \beq
        \langle RR'_1 \ DR_2 \rangle \ = \ \half \Nd\Nr(\Nrp-1)(\Nr-2)\Gp_1\Gp_2 + \Nd\Nr(\Nrp-1)\Gt_{12}
 ,\eeq
and we obtain that
 \beq
        \langle\gamma'_1\beta_2\rangle \ = \ \frac{2}{\Nr}\left[\frac{\Gt_{12}}{\Gp_1\Gp_2} - 1\right] 
        \ = \ 2\tur_{12} \,,
 \eeq
which is equal to the $\langle\gamma_1\beta_2\rangle$ of standard LS. 

Then the $\langle\gamma'_1\gamma'_2\rangle$:
 \beq
        \langle RR'_1\ RR'_2 \rangle \ = \ \sum_{\mu=1}^{\Ms}\sum_{\nu=1}^{\Ms}\langle R^\mu R^\mu_1\cdot R^\nu R^\nu_2\rangle \,,
 \eeq
where there are $\Ms(\Ms-1)$ terms with $\mu\neq\nu$ giving
 \beq
        \langle R^\mu R^\mu_1 \ R^\nu R^\nu_2\rangle \ = \ \langle R^\mu R^\mu_1\rangle\langle R^\nu R^\nu_2\rangle \ = \ \quarter (\Nrp)^2(\Nrp-1)^2\Gp_1\Gp_2
 ,\eeq
and $\Ms$ terms with $\mu=\nu$ giving
  \bear
        \langle R^\mu R^\mu_1\ R^\mu R^\mu_2\rangle & = & \quarter \Nrp(\Nrp-1)(\Nrp-2)(\Nrp-3)\Gp_1\Gp_2 + \nn\\
        & & + \Nrp(\Nrp-1)(\Nrp-2)\Gt_{12} \nn\\
        & & + \half \Nrp(\Nrp-1)\delta_{12}\Gp_1 \,.
 \eear
Adding these up gives
 \bear
        \langle RR'_1\ RR'_2 \rangle & = & \quarter \Nr(\Nrp-1)(\Nr \Nrp-\Nr-4\Nrp+6)\Gp_1\Gp_2 + \nn\\
        & & + \Nr(\Nrp-1)(\Nrp-2)\Gt_{12} + \nn\\
        & & + \half \Nr(\Nrp-1)\delta_{12}\Gp_1
 \eear
and 
 \bear
        \langle\gamma'_1\gamma'_2\rangle & = & \frac{2}{\Nr(\Nrp-1)}\left\{             
        2(\Nrp-2)\left[\frac{\Gt_{12}}{\Gp_1\Gp_2} - 1\right] + \farc{\delta_{12}}{\Gp_1} - 1\right\} \nn\\
        & = & 4\tur_{12} + \purp_{12} \,.
 \eear
This differs both from standard LS and from dilution, since it involves both $\Nr$ and $\Nrp$.

\end{document}